\documentclass[]{mnras}

\usepackage{url}
\usepackage{graphicx}
\usepackage{xspace}
\usepackage{amsfonts}
\usepackage{pifont}
\usepackage{textcomp}
\usepackage[hyphenbreaks]{breakurl}

\title[{\sl K2} observations of V729 Sgr]{V729 Sgr: A long period
dwarf nova showing negative superhumps during quiescence}

\author[]
{Gavin Ramsay$^{1}$, Matt A. Wood$^{2}$, John K. Cannizzo$^{3,4}$, Steve B. Howell$^{5}$, Alan Smale$^{6}$\\
  $^{1}$Armagh Observatory \& Planetarium, College Hill, Armagh, BT61 9DG, UK\\
  $^{2}$Department of Physics and Astronomy, Texas A\&M University--Commerce, Commerce, TX 75429, USA\\
   $^{3}$CRESST and Astroparticle Physics Laboratory NASA/GSFC, Greenbelt, MD 20771, USA\\
  $^{4}$Department of Physics, University of Maryland, Baltimore County, 1000 Hilltop Circle,
  Baltimore, MD 21250, USA\\
 $^{5}$NASA Ames Research Center, Moffett Field, CA 94095, USA\\
  $^{6}$NASA Goddard Space Flight Center, Greenbelt, MD 20771, USA}

\date{Accepted 2017 April 4. Received 2017 April 4; in original form 2017 February 17}

\begin{document}
\outer\def\gtae {$\buildrel {\lower3pt\hbox{$>$}} \over 
{\lower2pt\hbox{$\sim$}} $}
\outer\def\ltae {$\buildrel {\lower3pt\hbox{$<$}} \over 
{\lower2pt\hbox{$\sim$}} $}
\newcommand{\Msun}{$M_{\odot}$}
\newcommand{\lsun}{$L_{\odot}$}
\newcommand{\Rsun}{$R_{\odot}$}
\newcommand{\solar}{${\odot}$}
\newcommand{\kep}{\sl Kepler}
\newcommand{\ktwo}{\sl K2}
\newcommand{\swift}{\it Swift}
\newcommand{\Porb}{P_{\rm orb}}
\newcommand{\nuorb}{\nu_{\rm orb}}
\newcommand{\eplus}{\epsilon_+}
\newcommand{\eminus}{\epsilon_-}
\newcommand{\cd}{{\rm\ c\ d^{-1}}}
\newcommand{\MdotL}{\dot M_{\rm L1}}
\newcommand{\Mdot}{$\dot M$}
\newcommand{\Mdotsolar}{\dot{M_{\odot}} yr$^{-1}$}
\newcommand{\Ldisk}{L_{\rm disk}}
\newcommand{\src}{KIC 9202990}
\newcommand{\ergscm} {erg s$^{-1}$ cm$^{-2}$}
\newcommand{\rchi}{$\chi^{2}_{\nu}$}
\newcommand{\chisq}{$\chi^{2}$}
\newcommand{\pcmsq} {cm$^{-2}$}

\maketitle
\begin{abstract}
We report {\ktwo} observations of the eclipsing cataclysmic variable
V729 Sgr which covered nearly 80 days in duration. We find five short
outbursts and two long outbursts, one of which shows a clear plateau
phase in the rise to maximum brightness. The mean time between
successive short outbursts is $\sim$10 d while the time between the
two long outbursts is $\sim$38 d. The frequency of these outbursts are
unprecedented for a CV above the orbital period gap. We find evidence
that the mid-point of the eclipse occurs systematically earlier in
outburst than in quiescence. During five of the six quiescent epochs
we find evidence for a second photometric period which is roughly 5
percent shorter than the 4.16 h orbital period which we attribute to
negative superhumps. V729 Sgr is therefore one of the longest period
CVs to show negative superhumps during quiescence.
\end{abstract}


\section{Introduction}

After the second of the {\kep} satellite's reaction wheels failed in
May 2013, the space mission was re-purposed and renamed
{\ktwo}. Unlike the original mission which continuously monitored a
115 square degree field north of the Galactic plane between Cygnus and
Lyra for 4 years, {\ktwo} monitors fields along the ecliptic plane,
each pointing lasting $\sim$70--80 days (see Howell et
al. 2014). {\ktwo} has been used to study a wide variety of
astronomical sources including exoplanets, galaxies, solar system
objects, and cataclysmic variables (e.g., Dai et al. 2016; Kennedy et
al., 2016).

Briefly, cataclysmic variables (CVs) are interacting binaries in which
a low mass star (typically a K/M dwarf) fills its Roche lobe and the
material which flows through the L1 point forms an accretion disk
around the more massive white dwarf (if the white dwarf has a magnetic
field \gtae 1MG the disk is either partially or completely prevented
from forming). In the dwarf-novae sub-class of CVs, eruptions are seen
every few weeks or months where the brightness of the system increases
by $\sim$1-5 mag. These outbursts, arising from instabilities in the
accretion disk, are excellent sources with which to study the physics
of interacting binaries and accretion processes.

{\ktwo} observations during Campaign
9\footnote{https://keplerscience.arc.nasa.gov/k2-fields.html} took
place towards the end of 2015 and one of the sources observed was EPIC
214539533, the CV V729 Sgr. This star was first identified as being
variable in 1928 with further observations made by Ferwerda (1934)
showing it to be an irregular variable with successive maxima being
separated by 12--17 days. Although being fairly bright ($V\sim 14$
during outburst), V729 Sgr has not been particularly well studied,
apart from that of Cieslinski et al. (2000). These authors confirmed
its CV nature and found that V729 Sgr was an eclipsing system with a
period of 4.16 hrs.  In this paper, we report on the {\ktwo}
observations of V729 Sgr.

\begin{figure*}
\begin{center}
\setlength{\unitlength}{1cm}
\begin{picture}(12,9)
\put(-1.5,-1.5){\includegraphics{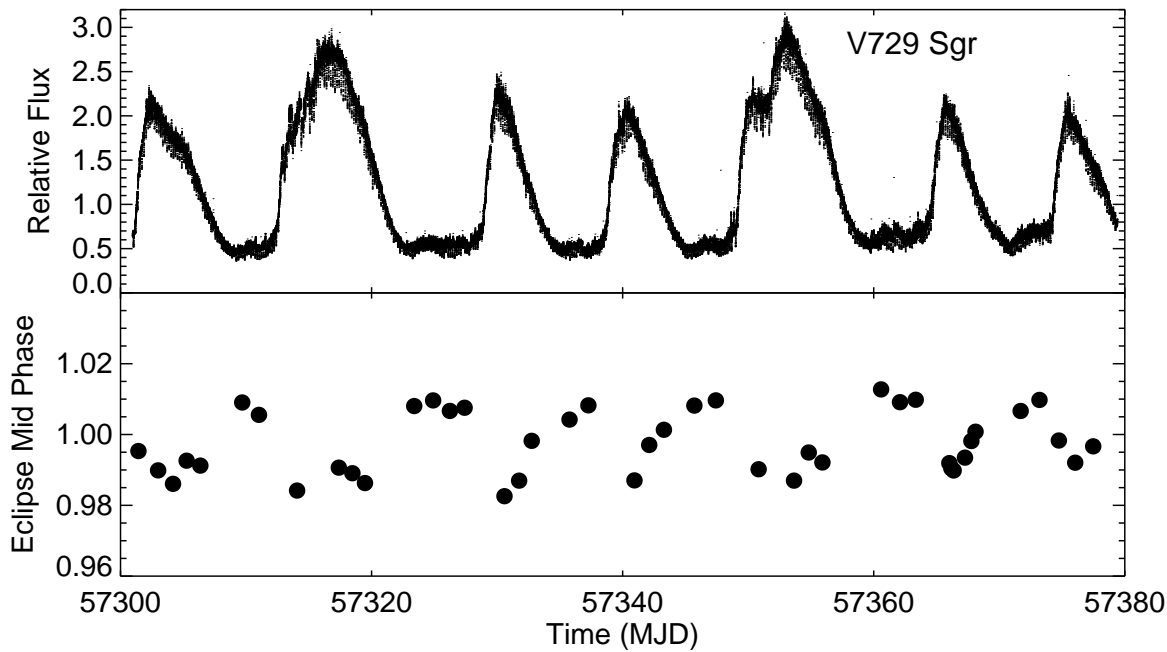}}
\end{picture}
\end{center}
\caption{Top Panel: The {\ktwo} short cadence light curve
  of V729 Sgr. Lower panel: the orbital phase of the eclipse where
  6--8 sucessive eclipses have been averaged to determine the eclipse mid-point.}
\label{light-phase} 
\end{figure*}

\section{Overview of the {\ktwo} observations}

{\ktwo} observations of V729 Sgr were made between MJD =
57301.0--57382.4 (2015 Oct 5 -- 2015 Dec 26) and were obtained in
Short Cadence (SC) mode which yields an effective exposure time of
58.8 sec. Since the {\ktwo} raw light curves of sources need to be
photometrically corrected for instrumental effects caused by minute
changes to its pointing (van Cleve et al., 2016), Andrew Vanderburg
very kindly supplied the corrected SC light curve using the techniques
outlined in Vanderburg \& Johnson (2014).

We show the light curve of V729 Sgr in the top panel of Figure
\ref{light-phase} and note that it is characteristic of a CV showing
regular outbursts. There are seven outbursts, two of which have higher
peak fluxes and significantly longer duration than the other
five: we call these `long' outbursts and the others `short'
  outbursts. On average there is an outburst every 12.7 d, with 
  short outbursts lasting $\sim$6.4 d and long lasting
$\sim$10.4 d. The mean time interval between successive short
outbursts is 10.0 d and the time interval between the two long
outbursts is $\sim$38 days. The orbital period of V729 Sgr (4.16 h) is
similar to that of the prototype dwarf nova U Gem (4.25 h). U Gem also
shows short and long outbursts with a mean outburst
recurrence time of $\sim$110 d (Szkody \& Mattei 1984), is much longer
than V729 Sgr.

\begin{figure}
\begin{center}
\setlength{\unitlength}{1cm}
\begin{picture}(6,7)
\put(-2,-7.5){\includegraphics{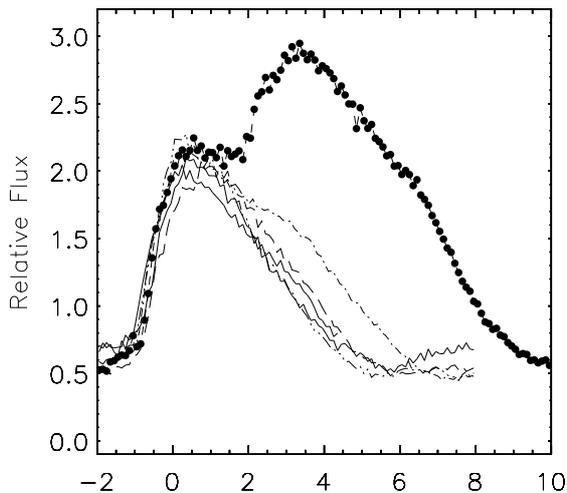}}
\end{picture}
\end{center}
\caption{The light curve (black dots) is the second long
  outburst observed for V729 Sgr using {\ktwo}. The superimposed lines
  show the five short outbursts where each peak time has been
  shifted (by eye) to the time zero, matching the start time of the
  planteau phase.  }
\label{SO-NO-Comp} 
\end{figure}

The second long outburst is shown in Figure \ref{SO-NO-Comp}.
After the initial rise from quiescence, the flux remains roughly
  constant for nearly 2 days (a feature which we call a `plateau'
  phase), after which there is a further increase in flux which is the
  long outburst. In Figure \ref{SO-NO-Comp} we show this 
  long outburst, superposed with the profiles of the five short
outbursts. We have shifted the short outbursts in time by-eye so
that each outburst profile has its `zero' point coinciding with the
peak of the outburst and the start of the plateau phase of the
  long outburst. We find that while there is a spread of $\sim$10
percent in the average short outburst amplitude, the amplitude
and rise time of the short outbursts are very similar to that of
the initial rise in the long outburst.

This behaviour is similar to the superoutbursts observed in the SU UMa
dwarf novae (which tend to have shorter orbital periods, $\la$2 h) and
the long outbursts seen in U Gem-type dwarf novae, both types of
systems which have been observed in detail using {\kep} (e.g. V447
Lyr, Ramsay et al. 2012). Indeed, it appears that all super and long
outbursts seen in dwarf nova have a short outburst precursor (see
Cannizzo 2012).  The first long outburst does not show such a clear
plateau phase as the second, (perhaps due to residual imperfections in
the corrections applied to the raw light curve), but appears to be
broadly consistent with such a view.

\section{The eclipse}

To investigate the light curve of V729 Sgr in more detail we extracted
sections during quiescence and fit each of these with a linear plus
Gaussian function. We then took the mid-point of each observed eclipse
(the times are given in Appendix 1) and determined a linear fit
to these times giving the ephemeris:

\begin{equation}
To = BMJD 57300.5183(5) + 0.173405(2)E
\end{equation}
%249.70 min, 4,16 h

\noindent
where the numbers in parentheses give the standard error on the last
digits. Our determined period, 4.1617 h, is consistent with that found
by Cieslinski et al. (2000) and we take this to be the orbital period
of V729 Sgr.

\begin{figure*}
\begin{center}
\setlength{\unitlength}{1cm}
\begin{picture}(12,9)
\put(-1.5,-1.5){\includegraphics{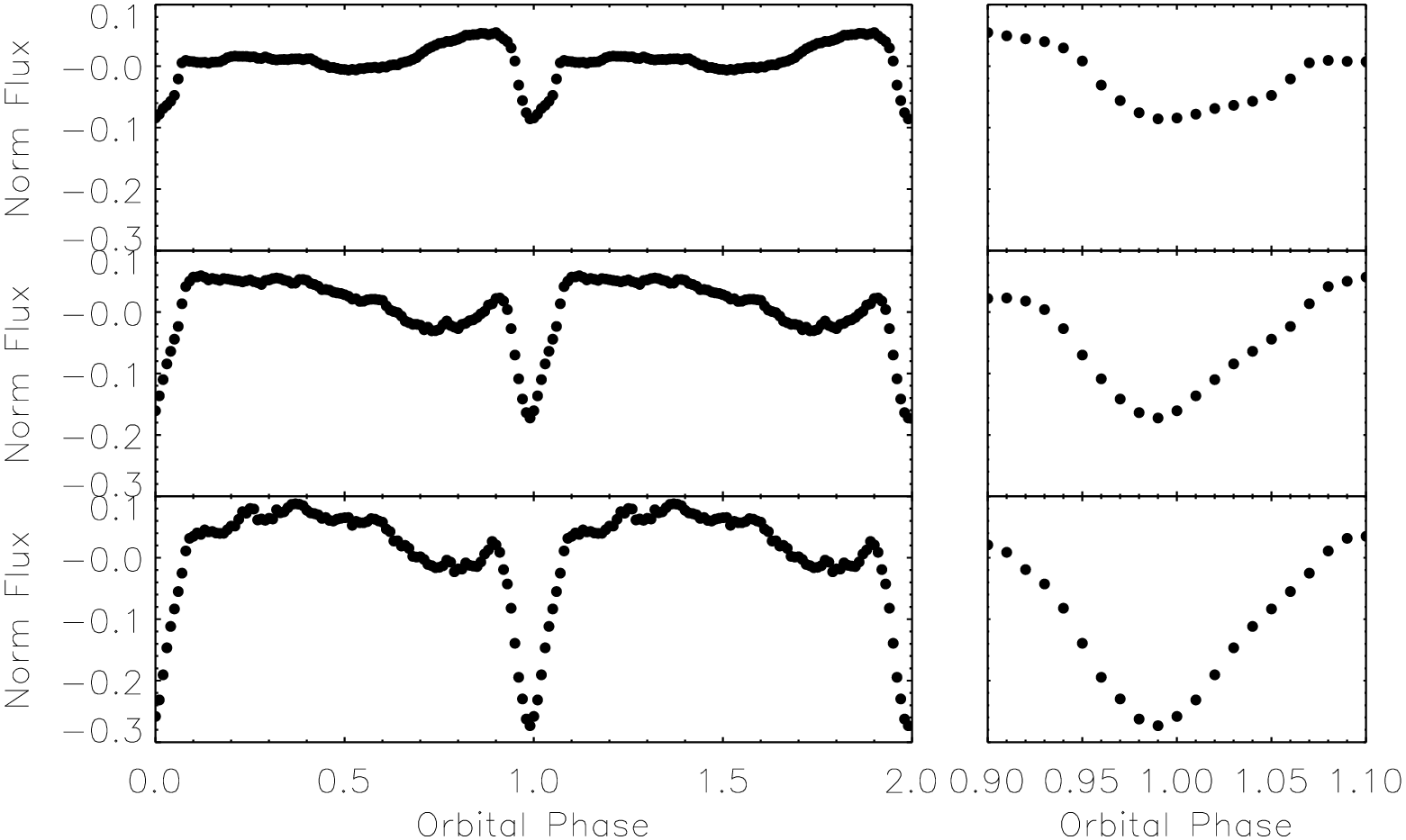}}
\end{picture}
\end{center}
\caption{The folded and binned light curves of V729 Sgr during
  quiescence (top panel); short outburst (middle panel) and 
    long outburst (bottom panel). In the right hand panels we
  highlight the profile of the eclipse.}
\label{fold} 
\end{figure*}

We then detrended the light curves covering the short and
long outbursts (i.e., the effects of the outbursts have been removed)
and determined the time of the eclipses from these data. Taking the
individual eclipse times and converting these to orbital phase using
equation 1, we find an indication that the mid-point of the eclipse
occurs earlier during outburst compared to quiescence. To reduce noise
in the eclipse profile, we formed the mean profile using 6--8 eclipses
(the exact number depended on how many eclipses there were between
outbursts) and determined the phase of the mid-point of the average
eclipse using a linear fit plus Gaussian profile. We show the
determined orbital phase of the eclipse over the course of the {\ktwo}
observations in the lower panel of Figure \ref{light-phase}. We find
that the mid-eclipse occurs at an earlier phase during outburst
compared to quiescence.

This finding is very similar to our result made using {\kep}
data of the U Gem dwarf nova V447 Lyr (Ramsay et al. 2012) and KIS
J192748.53+444724.5 (Scaringi, Groot \& Still 2013). During quiescence
a significant fraction of the light is emitted from the bright spot
where the accretion stream meets the accretion disk around the white
dwarf. The bright spot gets eclipsed {\sl after} the white dwarf has
been eclipsed. During outburst the contribution of the bright spot to
the system brightness is reduced and therefore during outburst the
mid-point of the eclipse appears {\sl earlier} compared to during
quiescence.

To explore the eclipse profile further, we took the detrended light
curves and folded and binned them to give mean orbital curves for
quiescence, short outbursts and long outburst and these
are shown in Figure \ref{fold}.  The quiescent light curve is very
similar to V447 Lyr and shows a peak at $\phi\sim$0.8--0.9 which is
due to the bright spot. During short and long outbursts
this bright peak, however, is not detected and instead a minimum is seen
at $\phi\sim$0.7--0.8. In addition, the eclipse becomes deeper from
quiescence to short outburst and into long outburst, which may
indicate that the system brightness is concentrated close to the white
dwarf (perhaps the boundary layer between the accretion disk and white
dwarf). The eclipse is also broader during outburst and indicates (as
was seen in V447 Lyr) that the disk has a greater extent compared to
quiescence.

\section{Superhumps}

CVs show a range a of periodic and quasi-periodic behaviour in their
light curves. In the absence of eclipses and/or phase resolved spectra
it can sometimes be difficult to determine the origin for periods
identified in power spectra. This is due to the fact that in addition
to there generally being a signature of the orbital period in the
light curve, CVs quite often show signatures of `superhumps'. The
`positive' superhumps are due to torsional disc oscillations and have
periods slightly longer than the orbital period (e.g., Wood et
al. 2011), whereas the `negative' superhumps result from a tilt and
retrograde precession of the accretion disc (e.g., Wood, Thomas, \&
Simpson 2009).  In the latter case, the accretion stream bright spot
transiting the face of the tilted disc provides the negative superhump
signal. For CVs with orbital periods shorter than $\sim$2 hrs,
positive superhumps are seen in almost all CVs found in superoutburst
(e.g., Patterson et al. 2005). The `period excess' over the orbital
period has been used to estimate the orbital period of many CVs.

There are a small number of CVs where negative superhumps have been
detected including during quiescence (e.g. Wood et al.\ 2011 reported
negative superhumps in the {\kep} observations of V344 Lyr, and Osaki
\& Kato 2013 reported the detection of negative superhumps over a full
supercycle using {\kep} observations of V1504 Cyg).  The physical
origin of the tilt of the disk negative superhumps has not been firmly
established, but Thomas \& Wood (2015) adapted the results of Lai
(1999) in smoothed particle hydrodynamic simulations to find that a
magnetic field on the primary can tilt the disk out of the orbital
plane.

\begin{figure}
\begin{center}
\setlength{\unitlength}{1cm}
\begin{picture}(6,14)
\put(-2,-1.0){\includegraphics{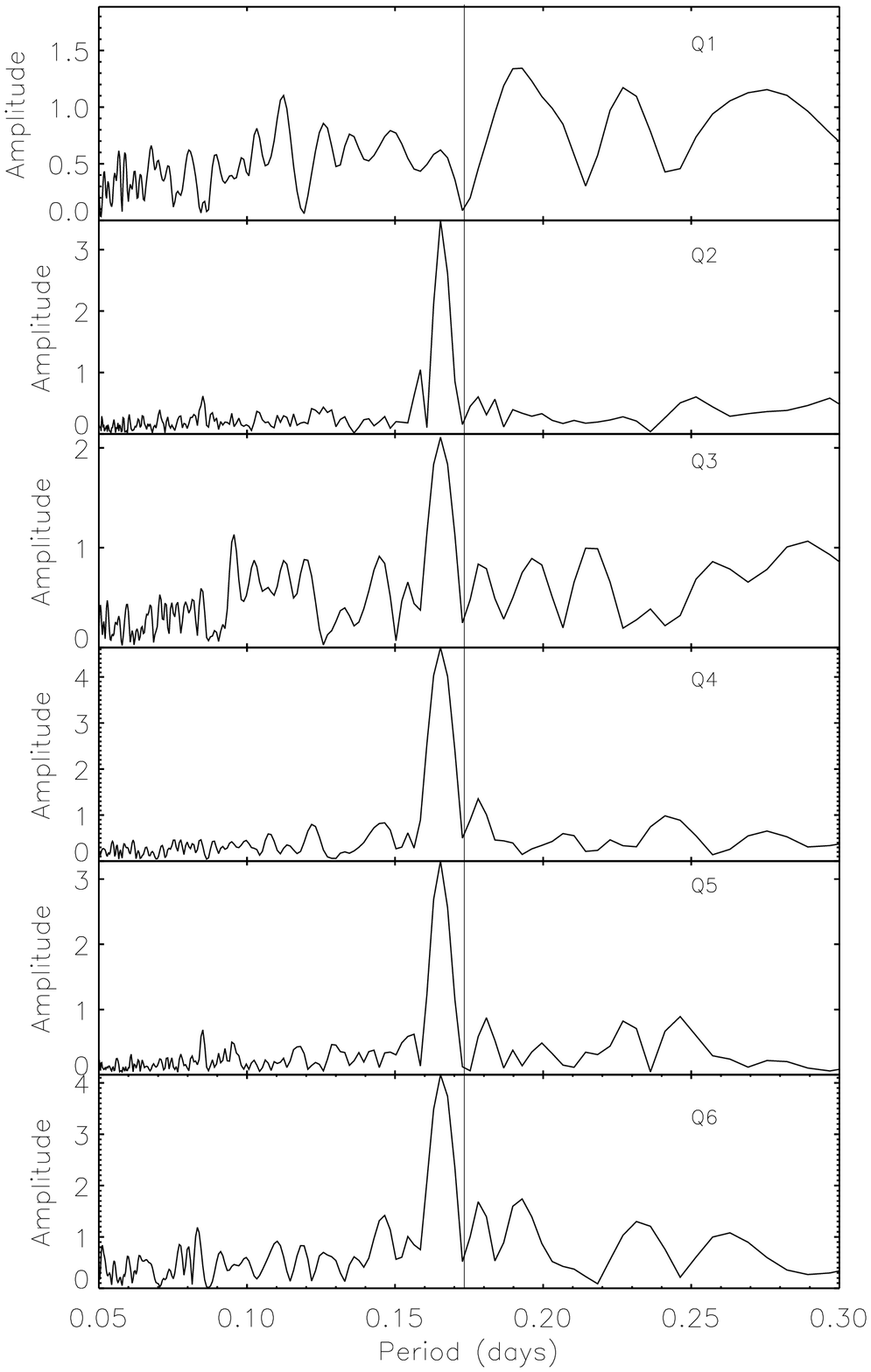}}
\end{picture}
\end{center}
\caption{We show the Discrete Fourier Transfor of the six quiescent
  light curves of V729 Sgr where we have first detrended the data and
  also pre-whitened the light curve using the orbital period and its
  three harmonics. The vertical line indicates the orbital period. The
  Q2--Q6 data indicate the presence of a negative super-hump.}
\label{power} 
\end{figure}

\begin{figure}
\begin{center}
\setlength{\unitlength}{1cm}
\begin{picture}(6,12.5)
\put(-2.2,-1.5){\includegraphics{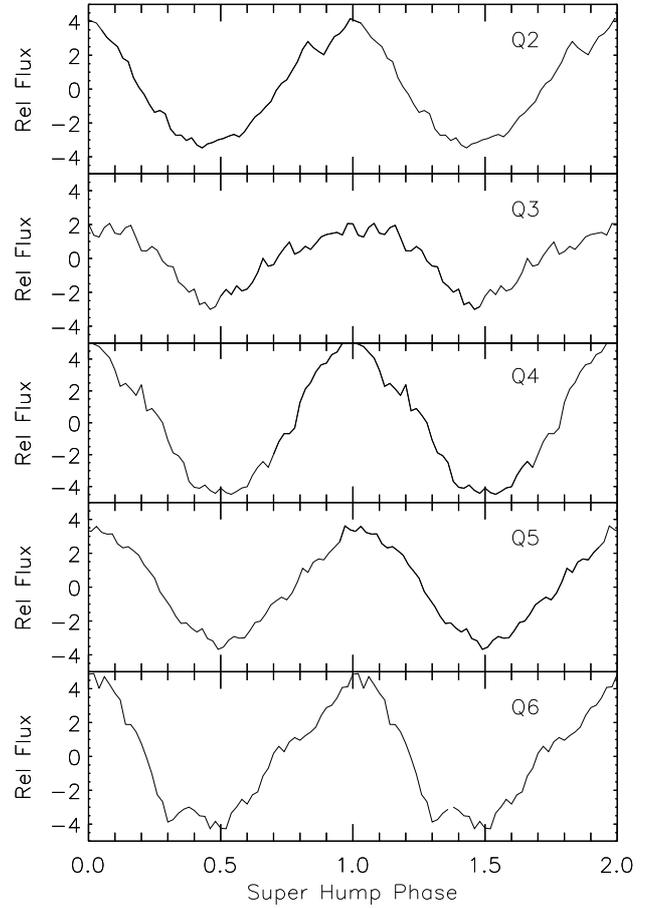}}
\end{picture}
\end{center}
\caption{For the five quiescent time intervals which show evidence of
  superhumps we folded and binned the detrended and prewhitend (on the
  orbital period and three harmonics) light curves on the period shown
  in Table \ref{excess}. We have phased the data so that flux maximum
  is defined as $\phi$=0.0 and we have plotted the folded light curve
  twice so that it runs from 0.0--2.0.}
\label{superfold} 
\end{figure}

We detrended the light curves for the short and long
outburst intervals seen in V729 Sgr and then obtained Lomb Scargle
Power Spectra and also Discrete Fourier Transforms (DFT) of these
light curves. The short or long outbursts show no evidence
for a period in the light curve other than $P_{orb}$ (which we know
precisely from the eclipses). However, we find evidence for a period
which is shorter than the orbital period in five out of the six
quiescent epochs. To investigate this in more detail we pre-whitened
the light curves by removing the orbital period and its three
harmonics and then obtained a DFT. We show the DFT of the six
detrended light curves in Figure \ref{power}. These DFT show clear
peaks at a period which is shorter than the orbital period in five of
the six quiescent light curves: we identify these peaks as a signature
of a negative superhump. We indicate the period of these peaks in
Table \ref{excess} and also the negative super-hump excess
$\epsilon^{-}=(P_{orb}-P_{sh})/P_{orb}$. We also did the same analysis
using a single light curve made up of the individual detrended Q2-Q6
quiescent light curves, and find a strong peak at 237.4 min. We folded
each of the quiescent light curves on the periods shown in Table
\ref{superfold}.  The mean negative excess, 0.049, is consistent with
the relationship between $\epsilon^{-}$ and $P_{orb}$, derived using
19 CVs in the study of Wood, Thomas \& Simpson (2009).

\begin{table}
\begin{center}
\begin{tabular}{llr}
\hline
Quiescence & $P_{sh}^{-}$ & $\epsilon^{-}$  \\
Section    & (min)       &                 \\
\hline
%Q1 & -      & -     &  275.90 & 0.105 \\
Q2 & 238.18 & 0.046 \\
Q3 & 236.30 & 0.054  \\
Q4 & 237.46 & 0.049  \\
Q5 & 237.60 & 0.048  \\
Q6 & 237.31 & 0.050  \\
\hline
\end{tabular}
\end{center}
\caption{We show the period identified in each of the six quiescent
  epochs and indicate whether they are likely negative superhumps
  ($P_{sh}^{-}$) or positive superhumps ($P_{sh}^{+}$) and note the
  positive $\epsilon^{+}$ and negative $\epsilon^{-}$ super-hump
  excess.}
\label{excess}
\end{table}

\section{Discussion}

\subsection{General Characteristics}

The upper panel of Fig. 1 indicates that V729 Sgr has
  characteristics which are typical of CVs, showing frequent
  short and long outbursts. The light curve also shows
eclipses, allowing the orbital period to be unambiguously determined,
(4.16 hr), placing it above the orbital period gap.  By determining
the phase of the mid-eclipse we find that the eclipse center occurs
earlier in phase during an outburst compared to quiescence (lower
panel of Fig 1). This is consistent with observations of other CVs,
and is due to the differing relative contribution of the bright spot
to the overall optical brightness in outburst compared to
quiescence. The duration of the short outbursts seen in V729 Sgr
are also typical of sources with similar orbital period, whilst
the duration of the long bursts are marginally shorter (we
  caution that we have a sample of only two long outbursts) than
other CVs with similar orbital period (see Fig 17 and 18 of
Otulakowska-Hypka et al. 2016).

\subsection{The frequency of outbursts}

\begin{figure}
\begin{center}
\setlength{\unitlength}{1cm}
\begin{picture}(6,6.3)
\put(-2.,-6.6){\includegraphics{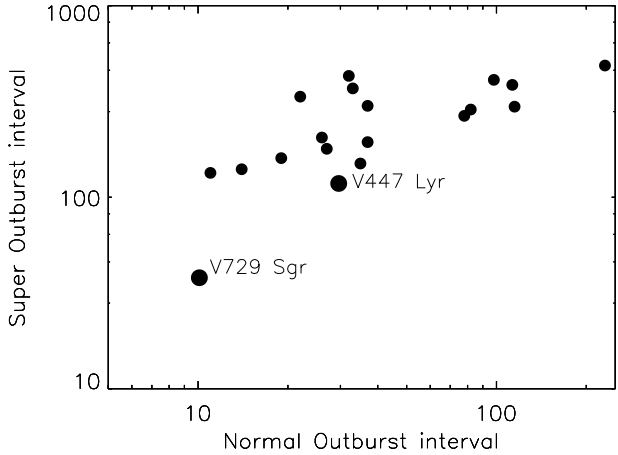}}
\end{picture}
\end{center}
\caption{We show the relationship between the outburst interval for
  short and super/long outbursts in CVs where we have taken
  the data from Ritter \& Kolb (2003). The source with the longest
  orbital period of 2.1 hrs. We have placed the location of V729 Sgr
  using the results from the K2 data.}
\label{rate-normal-super} 
\end{figure}

However, there are two features of the {\ktwo} light curve of V729 Sgr
which make this CV rather unusual. The first is the high frequency of
the outbursts, which is $\sim$10 d for successive short outbursts and
  38 d between the two long outbursts (which by necessity we assume is
  typical of the rate of long outbursts in this system). In Figure
  \ref{rate-normal-super} we plot the recurrence times of short and
  long outbursts for the 17 CVs with values reported in Ritter \& Kolb
  (2003) (all systems have an orbital period less than 2.1 hrs) and we
  add the values for V729 Sgr and V447 Lyr (Ramsay et al. 2012). The
  general trend, which was noted by Warner (1995), is that the short
  outburst rate is proportional to the super or long outburst rate.

The {\kep} observations of V447 Lyr (orbital period 3.74 hr) show it
is only a marginal outlier compared to the short period systems, but
V729 Sgr is clearly an outlier and falls into the lower left of Fig 2
of Warner (1995), which has only 3 systems in this part of the
distribution. These systems are the ER UMa systems (a sub-class of CVs
which also includes V1159 Ori and RZ LMi, see Kato et al. 2016) which
show outbursts every handful of days and superoutbursts every month or
so. This is unexpected (and so far not adequately explained) since a
short recurrence time is predicted to be due to a high mass accretion
rate. However, the known ER UMa CVs all have periods below the orbital
period gap at 2 hrs (RZ LMi has an orbital period of $\sim$83 min) and
are therefore expected to have relatively low mass transfer rates. In
the case of V729 Sgr we have a CV above the period gap (and hence
higher mass transfer rate) which shows frequent outbursts.

Dwarf novae exhibit a considerable scatter in outburst properties at a
given orbital period (Patterson 1984, Knigge et al. 2011), indicating
the mass transfer rate $\dot{M}_T$ feeding into the outer disk can
vary considerably from one system to another. This is presumably due
to the fact that the secondary stars can have widely varying histories
to end up at a given orbital period, as evidenced by the dispersion in
stellar radii at given orbital period (Knigge et al. 2011; see their
Figure 6).  In the disk instability model, higher $\dot{M}_T$
translates into a faster build-up of accretion disk material in
quiescence, and therefore more frequent outbursts. Therefore V729 Sgr
can be explained by invoking a higher-than-average $\dot{M}_T$.

Superoutbursts are usually defined as long outburst exhibiting
(positive) superhumps, and yet in V729 they are lacking. Our current
understanding for superhumps is that they are due to a torsionally
oscillating eccentric disk which undergoes prograde precession
(Whitehurst 1988). Physical conditions in the disk allow for this to
occur when the disk grows large enough so that the radius of 3:1
resonance with the binary orbital period is exceeded. Given that CV
primary masses are $\sim$1$M_{\odot}$ and the secondary mass is
$\sim$$0.1 M_{\odot} (P_{\rm orb}/{1 \ {\rm hr}})$, this means
superhumps can only occur in short orbital period systems $P_{\rm orb}
< 3$ hr. The mass ratio in V729 is probably not extreme enough to
support superhumps during outburst because the point of 3:1 resonance
lies beyond the last stable (i.e., non-intersecting) orbit.  Van
Paradijs (1983) carried out a study of long and short outbursts above
and below the $2-3$ hr period gap and concluded that 
  superoutbursts in the SU UMa systems are consistent in their
properties with long outbursts in systems above the period gap. In
other words, dwarf novae in general exhibit alternating series of long
and short outbursts. Systems below the period gap tend to have more
short outbursts between two long outbursts. The lack of superhumps in
the long outbursts in V729 Sgr and V447 Lyr indicates they are
not required for long outbursts to occur.
  
\subsection{The detection of negative superhumps}

The second feature of V729 Sgr which make it unusual is the presence
of negative superhumps during quiescence. Wood et al. (2009) list 21
CVs which have negative super-hump excesses reported in the
literature. The system with the longest orbital period was TV Col
which has a moderately magnetic white dwarf (and is an intermediate
polar). The next longest period system was LT Eri (SDSS J0407-0644)
with $P_{orb}$=4.08 hrs. The Ritter \& Kolb Catalogue 7.23 (2015 Dec)
(Ritter \& Kolb 2003) gives 27 sources which are reported as
displaying negative superhumps. Three have orbital periods longer than
V729 Sgr. AT Cnc ($P_{orb}$=4.83 h was reported to have negative
superhumps with $P_{-}$=4.65 h and 4.74 h made when it was in a bright
standstill episode (Kozhevnikov 2004). Quasi-Periodic signals were
seen in AY Psc ($P_{orb}$=5.22 h) over three years which were $\sim$5
percent shorter than $P_{orb}$ (G\"{u}lsecen et al. 2009) and
interpreted as negative superhumps (it is unclear whether these
superhumps were seen in quiescence or outburst). Finally, KIC 9406652
($P_{orb}$=6.11 h) was found to show a period of 5.75 hr (Gies et
al. 2013) which although present over the whole of the {\kep}
observations its amplitude varied over time, making KIC 9406652 the
longest period CV to show negative superhumps. V729 Sgr is therefore
unusual (although not unique) in showing negative superhumps above the
orbital period gap during quiescence.

\section{Conclusions}

V729 Sgr is worthy of further investigation since it shows outbursts
on a more frequent basis than its orbital period indicates. Future
observations should set out to identify further examples of
long outbursts and determine their frequency. Higher cadence
observations are encouraged at all accretion states to search for
superhumps (both positive and negative) to determine how often the
negative superhumps are observed in this system. It should shed
important light on the nature of accretion disks in high mass transfer
rate systems.

\section{Acknowledgments}

We thank Andrew Vanderburg for supplying us with the SC {\ktwo} light
curve of V729 Sgr.  Armagh Observatory is supported by the Northern
Ireland Government through the Dept for Communities.  This material is
based upon work supported by the National Science Foundation under
Grant No. AST-1305799 to Texas A\&M University-Commerce.

\appendix 

\section{Eclipse Times}

\begin{table}
%\centering
\caption{The time of the mid-eclipse (BMJD) is shown with the cycle number in the right hand column (see \S 3 for details).}
\label{eclipsetimes}
\begin{tabular}{lrlrlr}
\hline
Time   & Cycle & Time   & Cycle & Time & Cycle\\ 
(BMJD)      &  & (BMJD)      &  & (BMJD)  & \\
\hline
   57309.540  &      52  &57338.492   &    219 & 57375.430  &      432\\
   57309.707  &      53  &57338.668   &    220 & 57375.605  &      433\\
   57309.880  &      54  &57338.840   &    221 & 57375.777  &      434\\
   57310.050  &      55  &57339.016   &    222 & 57375.940  &      435\\
   57310.227  &      56  &57339.188   &    223 & 57376.130  &      436\\
   57310.406  &      57  &57339.360   &    224 & 57376.300  &      437\\
   57310.574  &      58  &57346.645   &    266 & 57376.470  &      438\\
   57310.746  &      59  &57346.816   &    267 & 57376.640  &      439\\
   57310.926  &      60  &57346.992   &    268 & & \\
   57311.094  &      61  &57347.164   &    269 & & \\
   57311.270  &      62  &57347.336   &    270 & & \\
   57311.445  &      63  &57347.510   &    271 & & \\
   57311.613  &      64  &57347.684   &    272 & & \\
   57311.790  &      65  &57347.855   &    273 & & \\
   57311.965  &      66  &57348.030   &    274 & & \\
   57312.137  &      67  &57348.203   &    275 & & \\
   57323.586  &     133  &57348.380   &    276 & & \\
   57323.750  &     134  &57348.550   &    277 & & \\
   57323.930  &     135  &57348.900   &    279 & & \\
   57324.100  &     136  &57349.070   &    280 & & \\
   57324.277  &     137  &57349.242   &    281 & & \\
   57324.453  &     138  &57349.420   &    282 & & \\
   57324.620  &     139  &57349.594   &    283 & & \\
   57324.797  &     140  &57349.770   &    284 & & \\
   57324.970  &     141  &57349.940   &    285 & & \\
   57325.140  &     142  &57362.074   &    355 & & \\
   57325.316  &     143  &57362.250   &    356 & & \\
   57325.490  &     144  &57362.420   &    357 & & \\
   57325.660  &     145  &57362.600   &    358 & & \\
   57325.840  &     146  &57362.777   &    359 & & \\
   57326.010  &     147  &57362.950   &    360 & & \\
   57326.180  &     148  &57363.117   &    361 & & \\
   57326.355  &     149  &57363.290   &    362 & & \\
   57326.527  &     150  &57363.465   &    363 & & \\
   57326.703  &     151  &57363.637   &    364 & & \\
   57326.875  &     152  &57363.812   &    365 & & \\
   57327.047  &     153  &57363.984   &    366 & & \\
   57327.223  &     154  &57364.156   &    367 & & \\
   57327.395  &     155  &57364.332   &    368 & & \\
   57327.570  &     156  &57364.504   &    369 & & \\
   57327.742  &     157  &57364.676   &    370 & & \\
   57327.914  &     158  &57364.850   &    371 & & \\
   57328.090  &     159  &57365.023   &    372 & & \\
   57328.260  &     160  &57365.195   &    373 & & \\
   57328.438  &     161  &57365.370   &    374 & & \\
   57328.610  &     162  &57365.543   &    375 & & \\
   57328.785  &     163  &57365.720   &    376 & & \\
   57328.957  &     164  &57365.890   &    377 & & \\
   57336.418  &     207  &57366.070   &    378 & & \\
   57336.586  &     208  &57366.242   &    379 & & \\
   57336.766  &     209  &57373.690   &    422 & & \\
   57336.934  &     210  &57373.867   &    423 & & \\
   57337.105  &     211  &57374.043   &    424 & & \\
   57337.280  &     212  &57374.215   &    425 & & \\
   57337.453  &     213  &57374.387   &    426 & & \\
   57337.625  &     214  &57374.562   &    427 & & \\
   57337.797  &     215  &57374.740   &    428 & & \\
   57337.977  &     216  &57374.906   &    429 & & \\
   57338.145  &     217  &57375.082   &    430 & & \\
   57338.320  &     218  &57375.258   &    431 & & \\
\hline   
\end{tabular}         
\end{table}           
                      
\end{document}